\documentclass[11pt]{article}
\usepackage[dvipdfm]{graphicx}
\usepackage{amsmath}
\begin{document}

\centerline{\bf Market-wide price co-movement around crashes in the Tokyo Stock Exchange}

\vspace{10pt}

\centerline{ {\bf Jun-ichi Maskawa}$^{\rm a}$\footnote{\label{q}Jun-ichi Maskawa, 
Department of Economics, Seijo University
Address: 6-1-20 Seijo, Setagaya-ku, Tokyo 157-8511, Japan
Tel.: +81-3-3482-5938, Fax: +81-3-3482-3660
E-mail: maskawa@seijo.ac.jp }
 ,  {\bf Joshin Murai}$^{\rm b}$, and {\bf Koji Kuroda}$^{\rm c}$}

\vspace{10pt}

\centerline{$^{\rm a}$Department of Economics, Seijo University} \vspace{10pt}

\centerline{$^{\rm b}$Graduate School of Humanities and Social Sciences, Okayama University}

\vspace{10pt}

\centerline{$^{\rm c}$Graduate School of Integrated Basic Sciences, Nihon University}
\vspace{10pt} \vspace{10pt}

\parindent=0pt \textbf{Abstract}

As described in this paper, we study market-wide price co-movements around crashes by analyzing a dataset of high-frequency stock returns of the constituent issues of Nikkei 225 Index listed on the Tokyo Stock Exchange for the three years during 2007--2009. Results of day-to-day principal component analysis of the time series sampled at the 1 min time interval during the continuous auction of the daytime reveal the long range up to a couple of months significant auto-correlation of the maximum eigenvalue of the correlation matrix, which express the intensity of market-wide co-movement of stock prices. It also strongly correlates with the open-to-close intraday return and daily return of Nikkei 225 Index. We also study the market mode, which is the first principal component corresponding to the maximum eigenvalue, in the framework of Multi-fractal random walk model. The parameter of the model estimated in a sliding time window, which describes the covariance of the logarithm of the stochastic volatility, grows before almost all large intraday price declines of less than -5\%. This phenomenon signifies the upwelling of the market-wide collective behavior before the crash, which might reflect a herding of market participants.

\vspace{10pt}

\parindent=0pt \textbf{JEL:}
C13,C22,G01

\parindent=0pt \textbf{Keywords:}
principal component analysis, market mode, multi-fractal random walk, precursor to market crash
%\begin{multicols}{2}

\newpage

\centerline{\bf Market-wide price co-movement around crashes in the Tokyo stock exchange}

\vspace{12pt}

\section{Introduction}
\parindent=22pt \parskip=0pt
Main stock markets throughout the world experienced considerable stock price declines time and time again during 2007--2009 because of ripple effects of the US sub-prime loan problem and the subsequent financial crisis. 

\begin{figure}[htbp]
\begin{center}
\includegraphics[bb=0 0 500 500,clip,width=12cm]{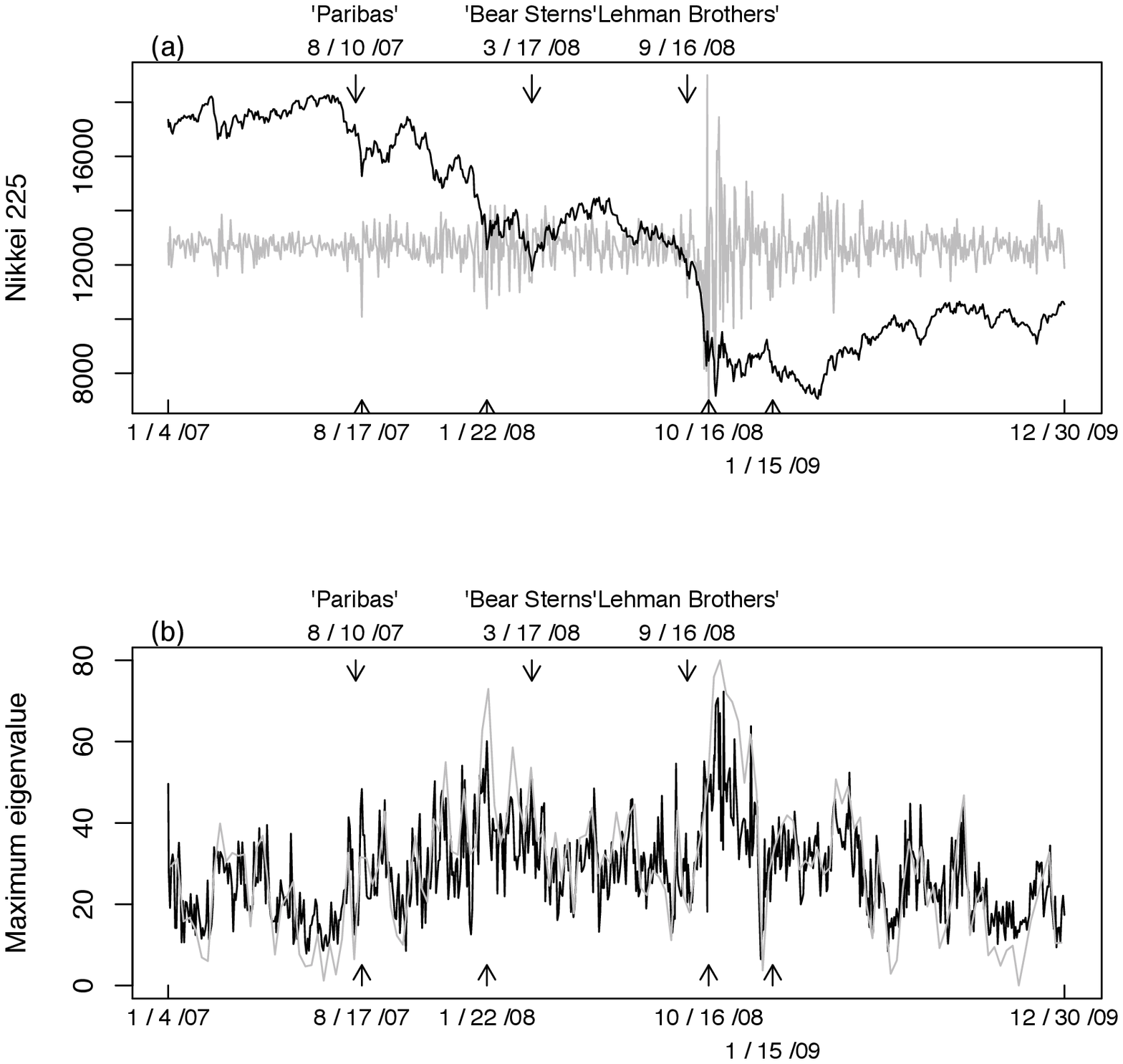}
\caption{Large price declines during January 2007 -- December 2009. (a) Time series of the Nikkei 225 Index (black line) and the open-to-close intraday log-return (gray line). (b) Time evolutions of the maximum eigenvalues of the cross-correlation matrices $\textbf{\textit{C}}$ for 1-min and 5-min log-return (black and gray lines respectively)}
\label{nikkei225}
\end{center}
\end{figure}

In Figure \ref{nikkei225}(a), the time series of Nikkei 225 Index and the open-to-close intraday log-return are shown for the period of January 2007 -- December 2009. The Nikkei 225 Index dropped during the period and, at the beginning of 2009, dropped below half of the value at January 2007. In 16 days among all trading days, a large price decline with a daily log-return (close to open overnight log-return + open-to-close intraday log-return) of less than -5\% was observed. Those are shown in Table \ref{large decline} in ascending order of the daily log-return. The date of 16 Sep 2008 is the next day of the bankruptcy of Lehman Brothers. Results show that the price drop was probably caused by the Lehman shock. The daily price change triggered by Paribas' shock on 9 Aug 2007, and the buyout of Bear Stearns by JP Morgan on 17 March 2008 are, respectively, -2.4\% and -3.8\%, which are not listed in the table. Those are not such large values because the standard deviation of daily log-returns for the period is 2.1\%.

However, 12 dates out of the whole of 16 dates listed in the table are concentrated within the two months after the market crash on 8 October 2008. Those heavy price falls occurred in a pessimistic mood for the world economy. Neither is related to specific news that might justify the magnitude of the subsequent drop-off (e.g. Cutler et al., 1989, Fair, 2002, Joulin et al., 2008). Instead, Maskawa (2012) pointed out that news including the words "financial crisis" came out every day. They did not impact stock prices severely alone, but they exacerbated the pessimistic mood prevailing among stock market participants. Such news increased after the Lehman shock preceding the market crash. The cases on 17 Aug 2007 and on 22 Jan 2008 have no specific shattering exogenous shocks. Consequently, they are thought to be derived endogenously.

 \begin{table}[htdp]
\caption{Daily log-return of less than -5\% during 4 Jan 2007 -- 30 Dec 2009 (733 trading days). Open-to-close intraday log-returns are also shown. The date of 9/16/08 is the day following the bankruptcy of Lehman Brothers. The standard deviation of daily log-returns for the period is 2.1\%.}
\begin{center}
\begin{tabular}{|c|c|c|c|c|c|c|c|}\hline

Rank&Date&Open&High&Low&Close&Daily~return&	Intraday~return\\ \hline
1&10/16/08&	9400.85&	9400.85&	8458.45&	8458.45&	-12.1\%&-10.6\%\\
2&10/10/08&	9016.34&	9016.34&	8115.41&	8276.43&	-10.1\%&-8.6\%\\
3&10/24/08&	8391.04&	8391.04&	7647.07&	7649.08&	-10.1\%&-9.3\%\\
4&10/8/08&	10011.64&10011.64&9159.81&9203.32&-9.8\%&-8.4\%\\
5&11/20/08&	8149.77&	8149.79&	7703.04&	7703.04&	-7.1\%&-5.6\%\\
6&10/22/08&	9198.14&	9198.14&	8674.69&	8674.69&	-7.0\%&-5.9\%\\
7&11/6/08&	9373.65&	9380.3&	8806.71&	8899.14&	-6.8\%&-5.2\%\\
8&10/27/08&	7568.36&	7878.97&	7141.27&	7162.9&	-6.6\%&-5.5\%\\
9&12/2/08&	8266.32&	8266.32&	7863.69&	7863.69&	-6.6\%&-5.0\%\\
10&1/22/08&	13125.23&13125.23&12572.68&12573.05&-5.8\%&-4.3\%\\
11&12/12/08&	8599.12&	8610.73&8087.99&	8235.87&	-5.7\%&-4.3\%\\
12&8/17/07&	16035.38&16062.59&15262.1&15273.68&-5.6\%&-4.9\%\\
13&11/13/08&	8564.47&	8564.47&	8148.3&	8238.64&	-5.4\%&-3.9\%\\
14&10/31/08&	8958.22&	9012.31&	8576.98&	8576.98&	-5.1\%&-4.3\%\\
15&9/16/08&	12028.45&12028.45&11551.4&11609.72&-5.1\%&-3.5\%\\
16&1/15/09&	8309.38&	8309.38&7997.73&8023.31&-5.0\%&-3.5\%\\ \hline
\end{tabular}
\end{center}
\label{large decline}
\end{table}

Many authors pointed out that the co-movement of stock markets or the correlation among stock returns increased before and after the major market crash such as ``Black Monday'' in 1987 (e.g. Meric and Meric, 1997, Harmon et al., 2011, Sandoval Jr. and Franca, 2012). In the earlier paper (Maskawa, 2012), the author analyzes the multivariate time-series of stock returns of the constituents of the FTSE100 listed on the London Stock Exchange for the period from May 2007 to January 2009 to study precursors to the global market crashes in 2008. He reported that a sharp rise in a measure of the collective behavior of stock prices was observed before the market crash. In this paper, for the same purpose, we study the collective behavior of stock prices around endogenous crashes by analyzing the portfolio consisting of high-frequency stock returns of the constituent issues of Nikkei 225 Index listed on the Tokyo Stock Exchange for the three years during 2007--2009. In the next section, we introduce a ``market mode'', defined as the first principal component corresponding to the maximum eigenvalue of the correlation matrix of the multivariate time-series of log-returns of the portfolio. As a result of the day-to-day principal component analysis of the time series sampled at the 1 min time interval during the continuous auction of the daytime, we find the long range auto-correlation of up to a couple of months of the maximum eigenvalue of the correlation matrix, which express the intensity of market-wide co-movement of stock prices. It also correlates with the open-to-close intraday return of Nikkei 225 Index. In section 4, we study the market mode in the framework of Multi-fractal random walk model. The parameter of the model estimated in sliding time window, which describes the covariance of the logarithm of the stochastic volatility, grows before almost all large intraday price declines of less than -5\%. This phenomenon signifies an upwelling of the market-wide collective behavior before the crash, which might reflect a herding of market participants.

\section{Market mode}

We analyze the multivariate time-series of stock returns of the constituents issues of Nikkei 225 Index at the present moment of 13 April 2013 listed on Tokyo Stock Exchange for the three years during 2007--2009\footnote{The constituents of Nikkei 225 Index are updated frequently. We selected $N=213$ issues that had been listed on the Tokyo Stock Exchange throughout the period.}. We exclude the overnight price changes and specifically examine the intraday evolutions of returns\footnote{To average out the normalized returns of selected issues, we skip about 30 min on an average immediately after the market opening, varying from day to day until all the issues get opening prices, except issues which recorded no price throughout the day.}. The total length of the time-series investigated here is $T=178,107$ min.
The return of the issue $i$ is the difference of the logarithms of the log-price
of $t$ min and $t-\delta t$ min:

\begin{equation}
\delta X_i(t)=\log(P_i(t))-\log(P_i(t-\delta t)).
\end{equation}

To remove the effect of the intraday pattern of market activity from the time-series of log-returns, $\delta X_i(t)$ is divided by the standard deviation of the corresponding time of day for each issue $i$, which is obtained from the whole of the time-series and which is designated by $\delta \tilde{X}_i(t)$ here.

The cross-correlation matrix of the time-series of log-returns of the period $T'$ measured in $\delta t$ as defined as

\begin{equation}
\textbf{\textit{C}}=\frac{1}{T'-1}\textbf{\textit{G}}\textbf{\textit{G}}^t,
\end{equation}

\begin{equation}\textbf{\textit{G}}=\left(
\begin{array}{cccc}
g_{11}  & g_{12}& \ldots &g_{1T'} \\ 
g_{21}  & g_{22}& \ldots &g_{2T'} \\ 
\vdots & \vdots & \ddots & \vdots \\
g_{N1}  & g_{N2}& \ldots &g_{NT'} \\ 
\end{array}
\right).
\end{equation}

\noindent
The component $g_{ij}\ (i=1,\cdots,N \ j=1,\cdots,T')$ of $\textbf{\textit{G}}$ is the normalization of de-seasonalized log-return $\delta \tilde{X}_i(j\delta t)$,

\begin{equation}
g_{ij}=\frac{\delta \tilde{X}_i(j\delta t)-\mu_i}{\sigma_i},
\end{equation}
where $\mu_i$ and $\sigma_i$ respectively denote the average and the standard deviation of $\tilde{\delta X_i(t)}$ for the corresponding period.

The first principal component corresponding to the maximum eigenvalue of the correlation matrix $\textbf{\textit{C}}$ is called the market mode because the weights of the mode are uniformly distributed among all issues of Nikkei 225 market-wide portfolio (Gopikrishnan et al., 2001). In principal component analysis, the variance of principal components expresses the contribution to all the variation of multivariate time-series. Consequently, the maximum eigenvalue, which is the variance of the first principal component, expresses the intensity of the market-wide collective behavior. 

Here, we evaluate the temporal behavior of maximum eigenvalue for two time interval $\delta t$. One is day-to-day evaluation for $\delta t=1$ min. The other is evaluation for $\delta t=5$min in sliding time window. We have about 242 min a day on an average. The width of the sliding time window is fixed at five times 242 min to keep the data length in line with day-to-day evaluation for $\delta t=1$ min. The results are presented in Figure \ref{nikkei225}(b).

\begin{figure}[htbp]
\begin{center}
\includegraphics[bb=0 0 500 200,clip,width=12cm]{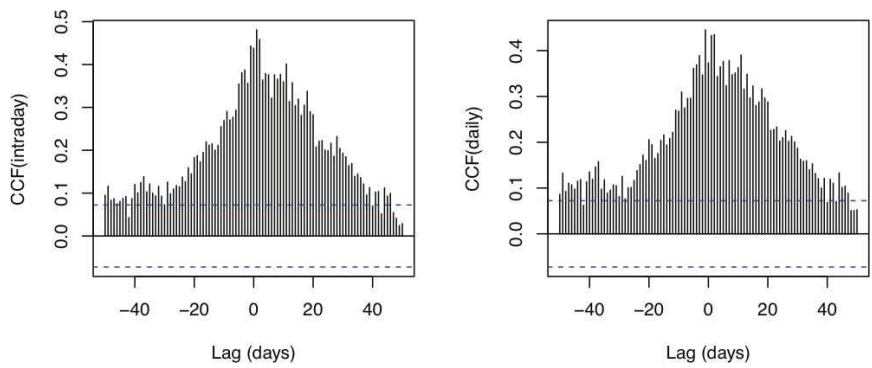}
\caption{Cross-correlation functions between the maximum eigenvalue for 1-min. log-return and absolute values of the open-to-close intraday log-return and daily log-return of the Nikkei 225 Index (left and right panels respectively).}
\label{ccf}
\end{center}
\end{figure}

\begin{figure}[htbp]
\begin{center}
\includegraphics[bb=0 0 500 500,clip,width=12cm]{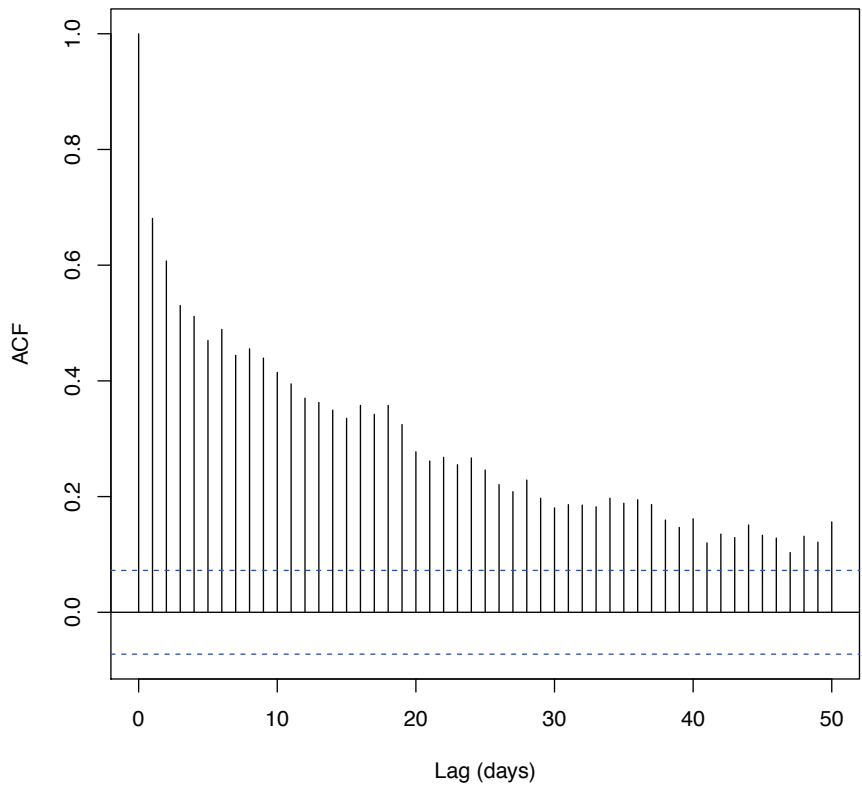}
\caption{Autocorrelation function of the maximum eigenvalue for 1-min. log-return.}
\label{acf}
\end{center}
\end{figure}

The maximum eigenvalue seems to increases when the stock price changes drastically. That indicates the intensification of the market-wide collective behavior.
The cross-correlation functions between the maximum eigenvalue and Nikkei 225 Index open-to-close intraday log-return and daily log-return is indeed statistically significant (Fig. \ref{ccf}). The maximum eigenvalue in the past (minus region of lag), even a couple of months, affects the current amplitude of log-return and vice versa. Moreover, the maximum eigenvalue has long memory of more than a couple of months (Fig. \ref{acf}).

\section{Multi-fractal random walk model}

Let $\delta M(t:\delta t)$ note the market mode for $\delta t$-min de-seasonalized log-returns. We use the framework of the Multifractal Random Walk (MRW) model (e.g. Muzy et al., 2000, Bacry et al., 2001) to analyze the time-series of the market mode $\delta M(t:\delta t)$.
We briefly introduce the MRW model in Appendix A, and show that the market mode is a multi-fractal random process and that it is well fitted by the model.

From equations (\ref{eq.cov}) and (\ref{eq.rho}), the variance of $\omega_{\delta t}$ the logarithm of the volatility is expressed by $\lambda^2\log(L/\delta t)$. We regard the variance $Var(\omega_{\delta t})$ as the quantity representing the intensity of the market-wide collective behavior, i.e. herding of market participants.

\begin{figure}[htbp]
\begin{center}
\includegraphics[bb=0 0 500 500,clip,width=12cm]{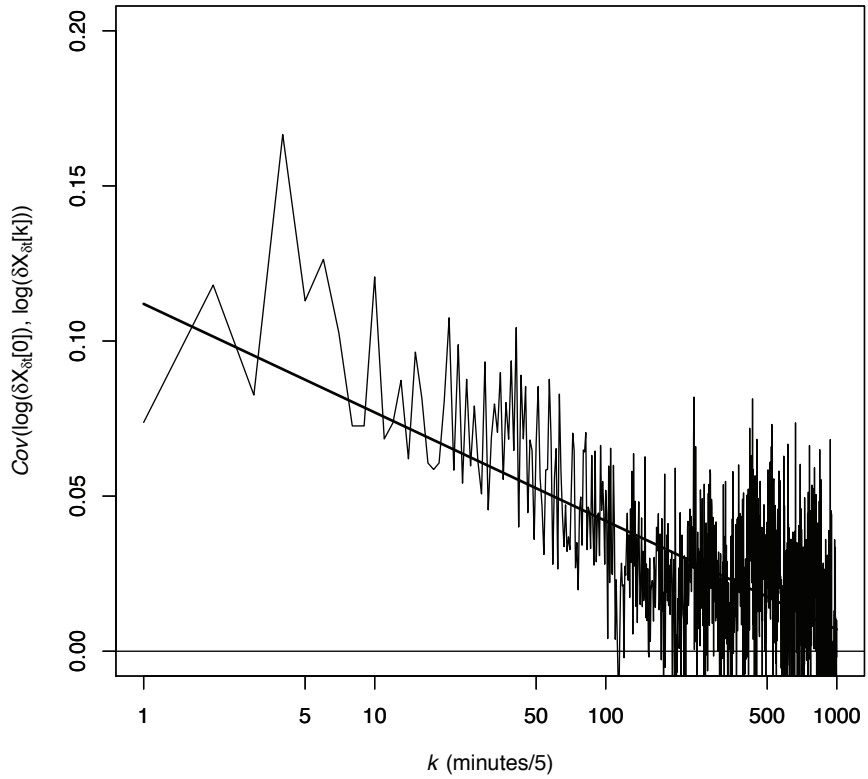}
\caption{Example of a covariance function for a time window. The semi-logarithmic plot of $Cov(\log(\delta X_{\delta t}[0]),\log(\delta X_{\delta t}[k]))$ against lag $k$ is shown. The straight line is the prediction of the MRW model (\ref{eq.covl}). The estimated values are $\lambda^2=0.014$ and $L/\delta t=1025.0$. $\delta t=5$ min}
\label{estimate}
\end{center}
\end{figure}

In the MRW model, $\omega_{\delta t}$ is a stationary process. Therefore, the parameters $\lambda$ and $L$ are constant against time. However, the intensity of the market-wide collective behavior is thought to undergo a change depending on the market phase. We therefore evaluate the temporal behavior of the variance $Var(\omega_{\delta t})$ in a sliding time window $[t-\Delta T,t ]$ with given width of $\Delta T$. 
To obtain the variance $Var(\omega_{\delta t})$ in each window, we must estimate parameters $\lambda$ and $L$. From analytical results of MRW model, the covariance of the logarithm of absolute return follows the equation (Bacry et al., 2001)

\begin{equation}
Cov(\log(\delta X_{\delta t}[i]),\log(\delta X_{\delta t}[i+k])) = 
\begin{cases}
-\lambda^2\log(|k|/L) & for~\delta t\ll |k|\le L \\
0 & for~L < |k|.
\end{cases}
\label{eq.covl}
\end{equation}
In Figure \ref{estimate}, an example of the covariance function for a time window is shown. The covariance function is fitted by the prediction of the MRW model (\ref{eq.covl}). Parameters $\lambda$ and $L$ are estimated using this equation. We set width $\Delta T=20,000$ min, which is in the range of $L < \Delta T < T$.
 
\begin{figure}[htbp]
\begin{center}
\includegraphics[bb=0 0 500 500,clip,width=12cm]{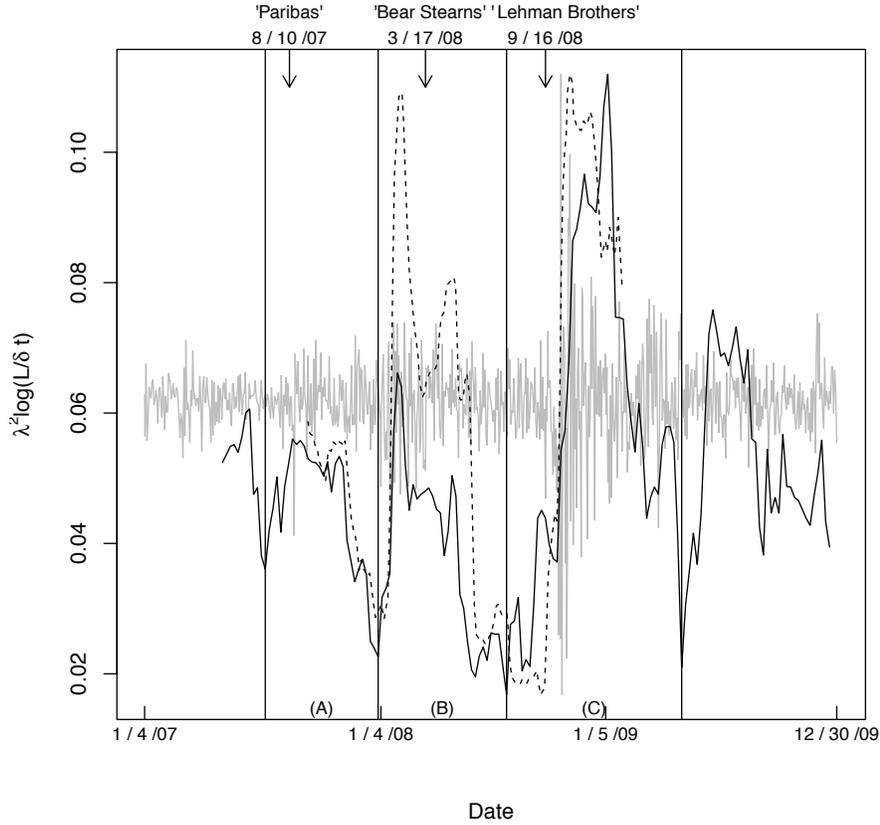}
\caption{Temporal evolution of the estimated variance $\lambda^2\log(L/\delta t)$for $\delta t = 5$ min (solid line). The same time axis shows the temporal evolution of the variance estimated by the market mode of London Stock Exchange for the period of May 2007 -- January 2009 shown in the earlier paper (Maskawa, 2012)(dashed line). The open-to-close intraday log-return of Nikkei 225 Index is also shown (gray line).}
\label{variance}
\end{center}
\end{figure}

Figure \ref{variance} portrays the temporal evolution of the variance $\lambda^2\log(L/\delta t)$ for $\delta t = 5$ min. The temporal evolution of the variance estimated by the market mode of London stock exchange on the same date is also shown. It is actually shown as lagged 9 hours behind time for the period from May 2007 to January 2009 shown in an earlier paper (Maskawa, 2012). Those two results for different stock markets coincide in up-and-down movement on a monthly time scale except minute change on smaller time scales. In both cases, the sharp rises of the variance $\lambda^2\log(L/\delta t)$, which mean the upwelling of the collective behavior of stock prices, are observed before the market crashes of Jan. and Oct. 2008. The same phenomena are observed before the large price decline of Aug. 2007 immediately after the Paribas' shock. The periods of high volatility lasted several months. However, steep declines of Mar. and Sep. 2008 are not observed, although they were thought to result respectively from the buyouts of Bear Stearns by J.P. Morgan on 17 Mar. and the Lehman shock on 15 September.

\begin{figure}[htbp]
\begin{center}
\includegraphics[bb=0 0 500 300,clip,width=12cm]{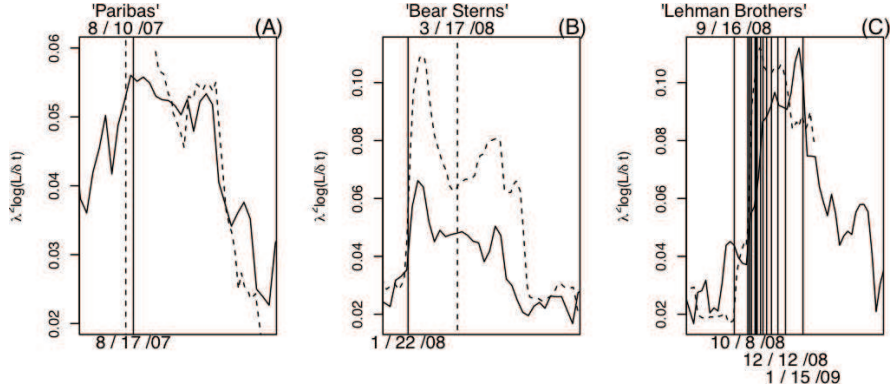}
\caption{Magnification of periods (A)--(C) marked in Figure \ref{variance}. The perpendicular lines of the dates of large price declines with a daily log-return of less than -5\% listed in Table \ref{large decline} are written (solid line). The lines for the Paribas' shock on 10 Aug. 2007 and the buyouts of Bear Stearns by J.P. Morgan on 17 Mar. 2008 are also written (dashed line). }
\label{variance_abc}
\end{center}
\end{figure}

A magnified image of the periods (A)--(C) marked in Figure \ref{variance} is shown in Figure \ref{variance_abc}. The perpendicular lines of the dates of large price declines with a daily log-return less than -5\% listed in Table \ref{large decline} are written in the figure (solid line). The lines for the Paribas' shock on 10 Aug. 2007 and the buyouts of Bear Stearns by J.P. Morgan on 17 Mar. 2008 are also written (dashed line). Fifteen of the sixteen dates listed in the table concentrate on the three ascending slopes of the variance $\lambda^2\log(L/\delta t)$\footnote{the price drop of 15 Jan. 2009 only occurred immediately after the local peak of the variance. }. This result indicates that the collective behavior of stock prices reflecting herd behavior of market participants is intensified by itself through the resulting price change.

\section{Conclusions}

We have shown through an empirical study of stock returns of the constituent issues of Nikkei 225 Index listed on Tokyo Stock Exchange for the period of Jan. 2007 -- Dec. 2009 that market-wide price co-movement becomes prominent before and after a large price decline such as endogenous market crash, which reflects intensification of the herd behavior of market participants.

From clustering of the maximum eigenvalue of market mode and the concentration of large price declines on the three ascending slopes of the variance of the logarithm of stochastic volatility of market mode in the period, herd behavior creates more intensified herd behavior through the resulting market price. Consequently, the price is self-fulfilling around the endogenous market crash.

\section*{Acknowledgements}

This research was partially supported by a Grant-in-Aid for Scientific Research (C) No. 24510203.

\appendix
\section{Multifractal Random Walk (MRW) model}

Multifractal Random Walk (MRW) $X(t)$ is a continuous process defined by the limit of the discrete random process $X_{\delta t}$:

\begin{equation}
X(t)=\lim_{\delta t \to 0,t=K_{\delta t}\delta t}X_{\delta t}(K_{\delta t}\delta t).
\end{equation}
The discrete process $X_{\delta t}(K_{\delta t}\delta t)$ is a standard random walk with a stochastic volatility that can be decomposed into subprocesses $\delta X_{\delta t}$:

\begin{equation}
X_{\delta t}(K_{\delta t}\delta t)=\sum_{i=1}^{K_{\delta t}}\delta X_{\delta t}[i].
\end{equation}
The subprocess $\delta X_{\delta t}(i)$ is described as

\begin{equation}
\delta X_{\delta t}[i]=\epsilon_{\delta t}[i]\exp({\omega_{\delta t}[i]}),
\end{equation}
where $\epsilon_{\delta t}$ is a stationary Gaussian white noise with variance $\sigma^2 \delta t$ and $\exp({\omega_{\delta t}[i]})$ is the stochastic volatility.

Bacry et al. show that  $X_{\delta t}(i)$ is a multifractal process (e.g. Muzy et al., 2000, Bacry et al., 2001), as

\begin{equation}
M(q,\delta t)=E(|\delta X_{\delta t}|^q)\sim \delta t^{\zeta_q}
\label{eq.mql}
\end{equation}
if $\omega_{\delta t}[i]$ is a stationary Gaussian process such that $E(\omega_{\delta t})=-Var(\omega_{\delta t})$ and

\begin{equation}
Cov(\omega_{\delta t}[i],\omega_{\delta t}[j])=\lambda^2 \log \rho _{\delta t}[|i-j|],
\label{eq.cov}
\end{equation}
where

\begin{equation}
\rho_{\delta t}[k] =
\begin{cases}
\frac{L}{(|k|+1)\delta t} & for~|k|\le L/\delta t -1\\
1 & otherwise.
\end{cases}
\label{eq.rho}
\end{equation}
Spectrum $\zeta_q$ is given by the formula

\begin{equation}
\zeta_q=(q-q(q-2)\lambda^2)/2.
\label{eq.zeta}
\end{equation}
To apply a MRW model to a time-series, three parameters $\sigma, \lambda$, and $L$ must be estimated.

\begin{figure}[htbp]
\begin{center}
\includegraphics[bb=0 0 500 500,clip,width=12cm]{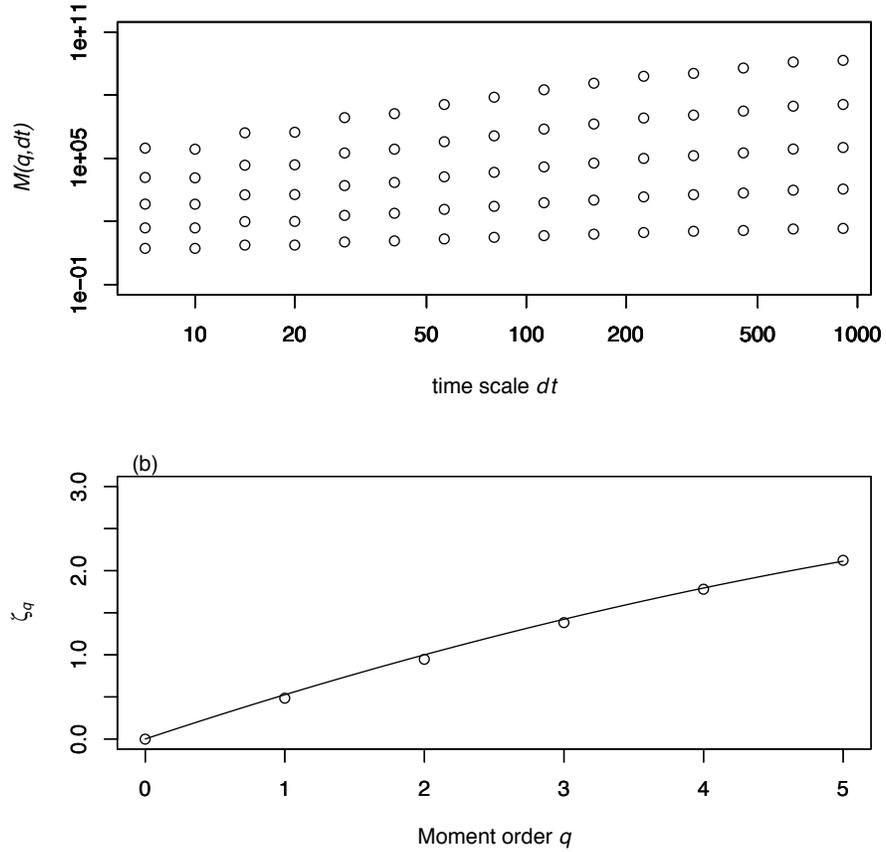}
\caption{Multifractal properties of the time-series of the market mode. (a) Double logarithmic plot of the expectation values of $q$ moments against the time scale $\delta t$ is shown. From top to bottom, the moment order $q=1, 2, 3, 4, 5$. Time scale $\delta t$ varies from 10 min to about 700 min (b) Spectrum $\zeta_q$ (circle) and the best fit by the theoretical prediction (\ref{eq.zeta}) are shown (solid line).}
\label{mql}
\end{center}
\end{figure}

Here, we check the applicability of MRW model to the market mode $\delta M(t:\delta t)$ using the formulae of (\ref{eq.mql}) and (\ref{eq.zeta}).
Figure \ref{mql} shows the double logarithmic plot of the expectation values of $q$ moments against the time scale $\delta t$, and the spectrum $\zeta_q$, which is estimated by eq. (\ref{eq.mql}). Those results mean that the MRW model is well applicable to the time series of the de-seasonalized market mode.

\end{document}